\documentclass[aps,prl,twocolumn,superscriptaddress]{revtex4}

\usepackage{graphicx,amsmath}
\begin{document}

\title{The Deformation of an Elastic Substrate by a Three-Phase Contact Line}
\author{Elizabeth R. Jerison}
\affiliation{Department of Physics, Yale University, New Haven, CT 06520}
\altaffiliation{Now at Department of Physics, Harvard University, Cambridge, MA 02138}
\author{Ye Xu}
\affiliation{Department of Mechanical Engineering and Materials Science, Yale University, New Haven, CT 06520}
\author{Larry A. Wilen}
\affiliation{Unilever Research and Development, Trumbull, CT 06611}
\author{Eric R. Dufresne}
\email[]{eric.dufresne@yale.edu}
\affiliation{Departments of Mechanical Engineering and Materials Science, Chemical and Environmental Engineering, Physics, and Cell Biology, Yale University, New Haven, CT 06520}

\date{\today}

\begin{abstract}
Young's classic analysis of the equilibrium of a three-phase contact line ignores the out-of-plane component of the liquid-vapor surface tension.
While it has long been appreciated that this unresolved force must be balanced by elastic deformation of the solid substrate, a definitive analysis has remained elusive because conventional idealizations of the substrate imply a divergence of stress at the contact line.
While a number of theories of have been presented to cut off the divergence, none of them have provided reasonable agreement with experimental data.
We measure surface and bulk deformation of a thin elastic film near a three-phase contact line using fluorescence confocal microscopy.
The out-of-plane deformation is well fit by a linear elastic theory incorporating an out-of-plane restoring force due to the surface tension of the gel.
This theory predicts that the deformation profile near the contact line is scale-free and independent of the substrate elastic modulus.
\end{abstract}


\maketitle

At first glance, it is hard to imagine that there is anything mysterious about a droplet of water resting on a solid surface.
However, mathematical descriptions of an idealized contact line, where the liquid, solid and vapor phases meet, can contain perplexing singularities.
For example, the diffusion-limited evaporation rate of a stationary droplet \cite{deegan.1997} and the strain-rate of a translating droplet \cite{hocking.1976, dussan.1991} both diverge at the contact line.

\begin{figure}[hbt]
\includegraphics[width=.9\columnwidth]{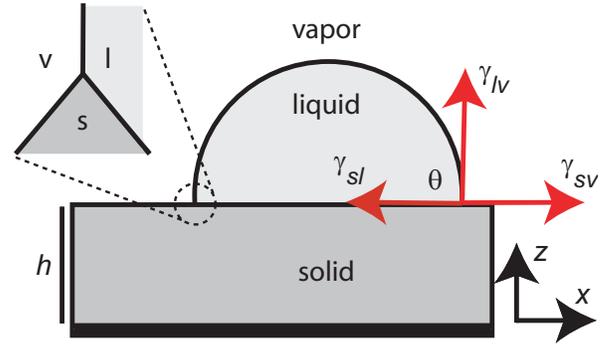}
\caption{\label{fig:schematic}  \emph{Schematic diagram showing a sessile droplet on an elastic substrate.}  Here, $\gamma_{sl}=\gamma_{sv}$.  Superimposed on the right side of the droplet is the conventional force balance used to derive Young's equation.   The anticipated scale-free deformation of the substrate at the contact line is superimposed on the left side.}
\end{figure}

As articulated by Young, the equilibrium angle of a three-phase contact line is determined by a balance
of solid-vapor, solid-liquid, and liquid-vapor surface tensions \cite{degennes}.
Surface tension is the derivative of the
free energy with respect to area and can be visualized as a  generalized force per unit length acting on the contact line, as shown in Fig. \ref{fig:schematic}.
Specifically, the
solid-liquid surface tension, $\gamma_{sl}$, exerts a force per unit length radially
inward, and the solid-vapor surface tension, $\gamma_{sv}$, exerts a force per unit
length radially outward.
The contact angle sets itself so that the
horizontal component of the liquid-vapor surface tension, $\cos(\theta) \gamma_{lv}$, balances $\gamma_{sl}-\gamma_{sv}$.
However, this classic force-balance leaves the vertical component of the liquid-vapor surface tension unbalanced.
While it is clear that the normal force due to the liquid-vapor surface tension must be balanced by elastic deformation of the solid substrate, the calculation of the deformation poses important conceptual difficulties.
In particular, the force per unit area, or stress, exerted by
the liquid-vapor surface tension diverges at the contact line.
Continuum elastic theory therefore predicts a strain divergence at the contact line.

To avoid this difficulty, Shanahan and DeGennes imposed a length-scale cut-off and focused their analysis on regions far from the contact line \cite{shanahan.1987,carre.1996}.
Alternatively, Long \emph{et. al.} \cite{long.1996} suggested that the divergence in the strain could be cut off by including a surface-tension penalty for the additional surface area of the deformed solid.
 Pericet-Camara \emph{et. al.} \cite{pericet.2009} recently measured the topography at the free surface of a PDMS substrate due to a drop of ionic liquid.
Their data showed  agreement with Long's theory in the long and short wavevector limits, but large discrepancies were found at distances from the contact line comparable to the substrate thickness.

\begin{figure*}[hbt]
\includegraphics[width=1.0\textwidth]{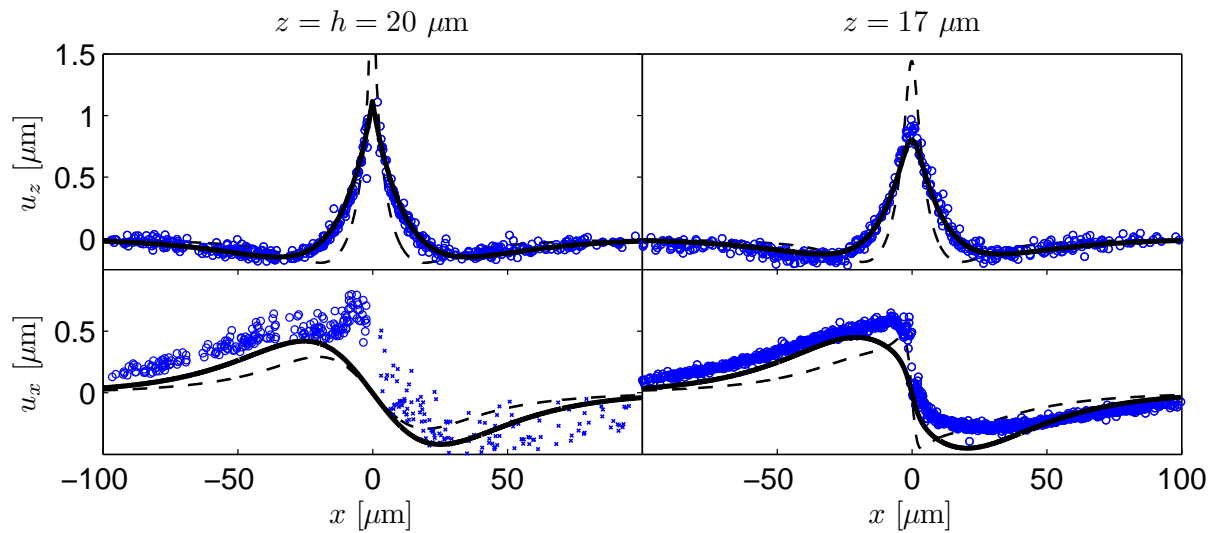}
\caption{\label{fig:measurement} \emph{Deformation of a thin elastic substrate due to a three-phase contact line.} Blue points indicate  deformations $u_z(x)$ (top) and $u_x(x)$ (bottom) measured at the surface ($z=h$, left) and just below the surface ($h-z=3$ $\mu$m, right) of a  gel near a three phase contact line.  The vapor phase is to the left of the peak and the liquid drop lies to the right of the peak.  Best fits to various theoretical models are superimposed on the data. The dashed lines are for a finite thickness elastic slab and the solid lines are for a finite thickness elastic slab with surface tension.  $x$-shaped plot symbols for $u_x(h)$ indicate points subject to systematic errors due to removal and re-deposition of beads by the contact line.  }
\end{figure*}

In this Letter, we report measurements of the in-plane and normal displacement
fields in a gel near its contact line with a sessile droplet of water.
We find that a linear elastic theory accounting for the finite thickness and surface tension of the gel resolves some of the conceptual difficulties imposed by the stress singularity and provides an excellent quantitative description of the out-of-plane deformation of the gel over all accessible length scales.
Interestingly, the deformation near the contact line is scale free and depends only on the ratio of the solid and liquid surface tensions.
While a careful accounting for surface tension gives a good description of the out-of-plane displacements, discrepancies remain for the in-plane displacements.

We measure the deformation of a highly-elastic silicone gel (Dow Corning Toray, CY52-276A/B) due to a 10 $\mu$L droplet of de-ionized water.
Uncured gel was spin-coated onto a relatively rigid glass coverslip to form a film of thickness $h=20$ $\mu$m.
We estimated the Young's modulus of the gel to be 3 kPa using bulk rheology.
To quantify the deformation field, two layers of fluorescent beads were incorporated
into the film, one at the free surface and one embedded 3 $\mu$m below the surface \cite{xu.2010}.
A spinning disc confocal microscope (Andor Revolution, mounted on a Nikon Ti Eclipse inverted microscope with an oil-immersion 40X objective, NA = 1.3) was used to image
the beads.
%OVER THE FIELD OF VIEW, THE CONTACT LINE DEVIATED FROM A STRAIGHT LINE BY AT MOST $6 \mu m$ DUE TO ITS CURVATURE.
Three-dimensional image stacks of the fluorescent tracer particles were acquired for a 5 minute period beginning about 15 minutes after the droplet was deposited on the surface.
Over this period, the contact line moved smoothly at approximately 36 $\mu$m/min as the drop evaporated.
  An image stack was also acquired about 45 minutes later to provide a zero-stress reference for bead positions.
Bead displacements were determined using centroid analysis and particle tracking software in MATLAB  \cite{crocker.1996, lu.2007}.
Since we expect forces to be invariant along the contact line, we average the bead displacements in this direction.
On the assumption that the force exerted by the contact line did not change significantly from one timestep to the next, we combined lab-frame displacement
profiles from four successive timesteps separated by 55 sec to construct a displacement profile as a function of distance from the contact line, as shown in shown in Fig. \ref{fig:measurement}.

The solid substrate forms a symmetric ridge about 1 $\mu$m high just below the contact line.
A wide and shallow valley appears on either side of the ridge, with resolvable deformations  observable up to 60 $\mu$m from the contact line.
In-plane displacements point toward the contact line and decay very slowly.
There is a slight but unmistakeable asymmetry in the $x$-displacements on either side of the contact line.
This  is surprising because the residual force of a static contact line on a smooth surface should be entirely out of plane, even if the contact angle is not identically 90$^\circ$.
We suspect that this small asymmetry is due to either pinning forces or viscous stresses near the slowly receding contact line.
As expected, the displacement fields are bounded, with no hint of a strain singularity near the contact line.

To construct a model that explains these observations, we begin with linear elastic theory. The linearized governing equation for the displacement field in an
isotropic elastic solid is:\cite{landau}\begin{equation}\label{eq:goveq}
(1-2\nu)\nabla^{2}\vec{u}+\nabla(\nabla\cdot\vec{u})=0,\end{equation}
 where $\nu$ is  Poisson's ratio ($\nu=\frac{1}{2}$ for incompressible
materials), and any forces are exerted at the boundaries.
Boussinesq
solved this equation in the case of an infinitely thick substrate \cite{landau}.
Integrating this solution along a line, we find the extensional strain, $\epsilon_{zz}$, at a depth $d$ directly below the contact line in an incompressible material,
 \begin{equation}
\epsilon_{zz}=\frac{3\gamma_{lv}}{2\pi E d},\end{equation}
where $\gamma_{lv}$ is the surface tension of the liquid as indicated in Fig. \ref{fig:schematic}, and $E$ is the Young's modulus of the substrate.
For simplicity, we assume a contact angle $\theta=\pi/2$ and that $d$ is much less than the length of the line.
Thus,
 strain diverges as $\epsilon_{zz} \sim \frac{1}{d}$ approaching the contact line.

\begin{figure}[hbt]
\includegraphics[width=.9\columnwidth]{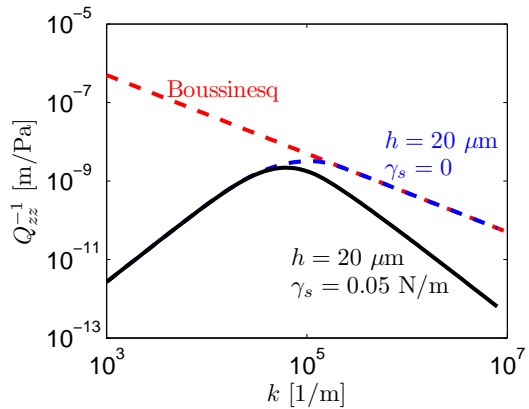}
\caption{\label{fig:compQ}\emph{Fourier transform of the deformation of an elastic substrate due to a out-of-plane line force.} The red dashed line shows the prediction of the classic theory of Boussinesq, which decays like $1/k$.  The blue dashed line supresses the divergence at $k=0$ by accounting for the finite thickness of the substrate.  The black line incorporates the surface tension of the substrate and decays as $1/k^2$ at large $k$.  }
\end{figure}

While Boussinesq's solution assumes a semi-infinite slab of elastic material, $h/R \gg 1$, our gel is much thinner than the radius of the drop, $h/R \approx 10^{-2}$.
We recently presented a solution to Eq. (\ref{eq:goveq}) for a finite-thickness
elastic film, which accounts for the zero-displacement boundary condition at the interface with a rigid substrate, located, in our coordinate system, at $z=0$ \cite{xu.2010}.
In Fourier space, the stresses at the surface, $\sigma_{iz}(h)$, are linearly related to displacements at height $z$, $u_j(z)$: \begin{equation}\label{eq:Q}
\sigma_{iz}(h)=Q_{ij}(z,h)u_{j}(z),\end{equation}
where the tensor $Q$ is a generalized spring constant and we sum over repeated indices.
For a three phase contact line, we exploit  translational invariance along the contact line and work in two dimensions.
Assuming that residual force from the contact line is normal to the surface, the stress due to the line can be written: $\sigma_{iz}(k)=\delta_{iz} \gamma_{lv}\sin{\theta}$.
Using the solution presented in Ref. \cite{xu.2010}, the $z$-displacements at the surface of an incompressible material due to the contact line are given by:
\begin{equation} \label{eq:thick}
Q^{-1}_{zz}(h,h)=\frac{3}{2E|k|} \left( \frac{e^{4|k|h}-4|k|he^{2|k|h}-1}{e^{4|k|h}+(4k^{2}h^{2}+2)e^{2|k|h}+1}\right),\end{equation}
Where $k$ is the wavenumber in the $x$-direction.
This solution reduces to the Boussinesq form in the limit of $h \rightarrow \infty$:
$Q_{zz}^{-1}=\frac{3}{2E}\frac{1}{|k|}$.
As shown in Fig. \ref{fig:compQ}, two salient features differentiate the
finite thickness solution (blue dashed line) from the Boussinesq solution (red dashed line).
First, the
substrate thickness $h$ introduces a length-scale so that $Q_{zz}^{-1}$ has a maximum at around $k=1/h$.
This leads to the dimples on either side of the peak.
Second,  the finite thickness solution suppresses a divergence in Boussinesq's solution as $|k|$ goes to zero.
This ensures that displacements go to zero as $x$ goes to $\pm \infty$.
However, the divergence at the origin remains.
Mathematically, the local behavior of the real-space
$z-$displacement at $x=0$ is governed by the inverse Fourier transform of
$\underset{|k|\rightarrow\infty}{lim}\,Q_{zz}^{-1}$.
In this
limit, the finite thickness solution and the Boussinesq solution are
equivalent: in both, $u_z$ diverges logarithmically, and $\epsilon_{zz}$ diverges like $1/z$.
Physically, the thickness is expected to be irrelevant in this region, since the solution should be dominated by local quantities near the singularity.

The measured displacement profiles were fit with this solution.
Since the film thickness, $h$, and the positions of the layers of beads, $z$, are known, this leaves one free fitting parameter, $\gamma_{lv}\sin{\theta}/E$.
We found the best-fit using least-squares minimization of $u_z$ for both layers of beads and $u_x$ for the inner layer of beads simultaneously.
Since the contact line removed and redeposited beads on the surface, we did not include $u_x$ from this layer in the fit.
The best fit is shown by the blue dashed line in Fig. \ref{fig:measurement}.
This theory not only fails to predict a bounded deformation at the surface, but it also  places the minima in $u_z$ too close the contact line, and systematically underestimates the in-plane deformations.

To improve this theoretical prediction,
 we take into account the surface tension of the gel and introduce a free energy penalty for the creation of new surface area on the substrate.
Following Long \cite{long.1996}, we continue to assume $\gamma_{sl}=\gamma_{sv} = \gamma_s$, which is valid for an equilibrium contact angle $\theta=\pi/2$.  The measured equilibrium contact angle was about 105$^{\circ}$, so this approximation is expected to be reasonable but not completely accurate.

Since the surface tension of the solid exerts a force per unit length like the surface tension of the liquid,
it is capable
of cutting off the singularity at the origin.
While Eq. (\ref{eq:Q})  still governs displacements within the bulk,  the stress at the surface now includes an internal contribution $\sigma_{\gamma}$ from the substrate surface tension, as well as the term due directly to the contact line, $\sigma_{ext}$:

\begin{equation}
\sigma_{ext,iz}=Q_{ij}(z,h)u_{j}(z)-\sigma_{\gamma,iz}.\end{equation}
 To arrive at the form of the surface stress, we first calculate the  free energy per unit length due to additional surface area, $\frac{1}{4}u_{z}^{2}k^{2}\gamma_{s}$.
 The associated stress is the derivative of this free energy with respect
to small changes in the $z-$displacement of the free surface: $\sigma_{\gamma,iz}=- \delta_{iz} k^{2}\gamma_{s}u_{z}(h)/2$, \cite{long.1996}.
Thus, we can  write:
 \begin{equation}\sigma_{ext,iz}=QS_{ij}(z,h) u_j(z),\end{equation} where
 \begin{equation}QS_{i,j}(z,h)=Q_{ij}(z,h)+\frac{\delta_{zi}k^2 \gamma_s}{2} Q^{-1}_{ik}(h,h) Q_{kj}(z,h)\end{equation}
To determine the $z-$displacements
in response to the line force, we examine $QS_{zz}^{-1}$.
Again, for the specific case of
$z=h$ and $\nu=\frac{1}{2}$:\begin{widetext}\begin{equation}
QS_{zz}^{-1}(h,h)=\frac{3(e^{4|k|h}-4|k|he^{2|k|h}-1)}{2E|k|(e^{4|k|h}+(4k^{2}h^{2}+2)e^{2|k|h}+1)-\frac{3}{2}k^2\gamma_{s}(-e^{4|k|h}+4|k|he^{2|k|h}+1)}.
\end{equation}\end{widetext}
This result, which incorporates the solid surface tension into the finite-thickness solution, is plotted as a solid black curve in Fig. \ref{fig:compQ}.
While it agrees with Eq. (\ref{eq:thick}) for small wavenumbers, it decays much faster at large wavenumbers, $QS_{zz}^{-1} \sim 1/|k|^{2}$.
Importantly, this fast decay ensures that $u_z$ and $\epsilon_{zz}$ remain finite at the contact line.
%The inverse Fourier transform of $\underset{k\rightarrow\infty}{\lim}u_z=\frac{-2 \gamma_{lv}}{\gamma_{s}k^{2}}$ is $u_{0}e^{-|x|/L}$.
Here, the shape of the surface is a cusp given by $u_z(0)-u_z(x) = \gamma_{lv} \sin{\theta} |x|/\gamma_s$.
This self-similar shape applies in the regime $|k|\gg E/\gamma_s$: at these small length scales, the elastic modulus does not effect the shape of the cusp.
Finally, the solid surface tension pushes the peak to lower $k$, increasing
the characteristic wavelength of the displacement profile.

This solution provides a much closer fit to the data.
The best-fit curves are shown as solid black lines in Fig. \ref{fig:measurement}.
The values extracted for the slope of the cusp $\gamma_{lv}\sin{\theta}/\gamma_{s}=0.14$ and the scale of out-of-plane deformation $\gamma_{lv}\sin{\theta}/E=3.2$ $\mu$m.
With the addition of the solid surface tension, our theory beautifully captures the out-of-plane deformation, including the shape of the peak and locations of minima.
However,  these values do not accurately fit the in-plane displacements.
This discrepancy may be due to in-plane forces due to pinning or viscous drag near the contact line.

In conclusion, we measure the displacement field in a linear, elastic substrate due
to the contact line of a water drop, and show that the $z-$displacement
profiles are well-fit by a model that includes both the finite thickness
of the substrate and the substrate's surface tension.
This model solves many of the conceptual problems raised by applying the Boussinesq solution to
the unresolved force of an equilibrium contact line.
The finite thickness of the substrate ensures
that displacements go to zero far
from the contact line.
The substrate's surface tension counters
the stress singularity at the origin, ensuring that the $z-$displacement and strain
at the origin do not diverge.
Additionally, work is needed to treat cases where the equilibrium contact angle is far from 90$^\circ$, indicating a large discrepancy in surface energies, $\gamma_{sl}$ and $\gamma_{sv}$, which we have taken to be identical.
This work enables a closer look at pinning and viscous dissipation near the contact line and provides a basis for the extension of traction force microscopy.
We acknowledge support from Yale College, Unilever and NSF DBI-0619674.

%\bibliography{refs}

\begin{thebibliography}{12}
\expandafter\ifx\csname natexlab\endcsname\relax\def\natexlab#1{#1}\fi
\expandafter\ifx\csname bibnamefont\endcsname\relax
  \def\bibnamefont#1{#1}\fi
\expandafter\ifx\csname bibfnamefont\endcsname\relax
  \def\bibfnamefont#1{#1}\fi
\expandafter\ifx\csname citenamefont\endcsname\relax
  \def\citenamefont#1{#1}\fi
\expandafter\ifx\csname url\endcsname\relax
  \def\url#1{\texttt{#1}}\fi
\expandafter\ifx\csname urlprefix\endcsname\relax\def\urlprefix{URL }\fi
\providecommand{\bibinfo}[2]{#2}
\providecommand{\eprint}[2][]{\url{#2}}

\bibitem[{\citenamefont{Deegan et~al.}(1997)\citenamefont{Deegan, Bakajin,
  Dupont, Huber, Nagel, and Witten}}]{deegan.1997}
\bibinfo{author}{\bibfnamefont{R.~D.} \bibnamefont{Deegan}},
  \bibinfo{author}{\bibfnamefont{O.}~\bibnamefont{Bakajin}},
  \bibinfo{author}{\bibfnamefont{T.~F.} \bibnamefont{Dupont}},
  \bibinfo{author}{\bibfnamefont{G.}~\bibnamefont{Huber}},
  \bibinfo{author}{\bibfnamefont{S.~R.} \bibnamefont{Nagel}}, \bibnamefont{and}
  \bibinfo{author}{\bibfnamefont{T.~A.} \bibnamefont{Witten}},
  \bibinfo{journal}{Nature} \textbf{\bibinfo{volume}{389}},
  \bibinfo{pages}{827} (\bibinfo{year}{1997}).

\bibitem[{\citenamefont{Hocking}(1976)}]{hocking.1976}
\bibinfo{author}{\bibfnamefont{L.~M.} \bibnamefont{Hocking}},
  \bibinfo{journal}{J. Fluid. Mech.} \textbf{\bibinfo{volume}{76}},
  \bibinfo{pages}{801} (\bibinfo{year}{1976}).

\bibitem[{\citenamefont{Dussan et~al.}(1991)\citenamefont{Dussan, Ram\'e, and
  Garoff}}]{dussan.1991}
\bibinfo{author}{\bibfnamefont{E.~B.} \bibnamefont{Dussan}},
  \bibinfo{author}{\bibfnamefont{E.}~\bibnamefont{Ram\'e}}, \bibnamefont{and}
  \bibinfo{author}{\bibfnamefont{S.}~\bibnamefont{Garoff}},
  \bibinfo{journal}{J. Fluid. Mech.} \textbf{\bibinfo{volume}{230}},
  \bibinfo{pages}{97} (\bibinfo{year}{1991}).

\bibitem[{\citenamefont{de~Gennes et~al.}(2004)\citenamefont{de~Gennes,
  Brochard-Wyart, and Quere}}]{degennes}
\bibinfo{author}{\bibfnamefont{P.-G.} \bibnamefont{de~Gennes}},
  \bibinfo{author}{\bibfnamefont{F.}~\bibnamefont{Brochard-Wyart}},
  \bibnamefont{and} \bibinfo{author}{\bibfnamefont{D.}~\bibnamefont{Quere}},
  \emph{\bibinfo{title}{Capillarity and wetting phenomena: drops, bubbles,
  pearls, waves}} (\bibinfo{publisher}{Springer}, \bibinfo{address}{New York},
  \bibinfo{year}{2004}).

\bibitem[{\citenamefont{Shanahan and de~Gennes}(1987)}]{shanahan.1987}
\bibinfo{author}{\bibfnamefont{M.~E.~R.} \bibnamefont{Shanahan}}
  \bibnamefont{and} \bibinfo{author}{\bibfnamefont{P.~G.}
  \bibnamefont{de~Gennes}}, \emph{\bibinfo{title}{Adhesion 11}}
  (\bibinfo{publisher}{Elsevier Applied Science}, \bibinfo{address}{London},
  \bibinfo{year}{1987}).

\bibitem[{\citenamefont{Carre et~al.}(1996)\citenamefont{Carre, Gastel, and
  Shanahan}}]{carre.1996}
\bibinfo{author}{\bibfnamefont{A.}~\bibnamefont{Carre}},
  \bibinfo{author}{\bibfnamefont{J.-C.} \bibnamefont{Gastel}},
  \bibnamefont{and} \bibinfo{author}{\bibfnamefont{M.~E.~R.}
  \bibnamefont{Shanahan}}, \bibinfo{journal}{Nature}
  \textbf{\bibinfo{volume}{379}}, \bibinfo{pages}{432} (\bibinfo{year}{1996}).

\bibitem[{\citenamefont{Long et~al.}(1996)\citenamefont{Long, Ajdari, and
  Leibler}}]{long.1996}
\bibinfo{author}{\bibfnamefont{D.}~\bibnamefont{Long}},
  \bibinfo{author}{\bibfnamefont{A.}~\bibnamefont{Ajdari}}, \bibnamefont{and}
  \bibinfo{author}{\bibfnamefont{L.}~\bibnamefont{Leibler}},
  \bibinfo{journal}{Langmuir} \textbf{\bibinfo{volume}{12}},
  \bibinfo{pages}{5221} (\bibinfo{year}{1996}).

\bibitem[{\citenamefont{Pericet-Camara
  et~al.}(2009)\citenamefont{Pericet-Camara, Auernhammer, Koynov, Larenzoni,
  and Bonaccurso}}]{pericet.2009}
\bibinfo{author}{\bibfnamefont{R.}~\bibnamefont{Pericet-Camara}},
  \bibinfo{author}{\bibfnamefont{G.~K.} \bibnamefont{Auernhammer}},
  \bibinfo{author}{\bibfnamefont{K.}~\bibnamefont{Koynov}},
  \bibinfo{author}{\bibfnamefont{R.}~\bibnamefont{Larenzoni},
  \bibfnamefont{S.~adn~Raiteri}}, \bibnamefont{and}
  \bibinfo{author}{\bibfnamefont{E.}~\bibnamefont{Bonaccurso}},
  \bibinfo{journal}{Soft Matter} \textbf{\bibinfo{volume}{5}},
  \bibinfo{pages}{3611} (\bibinfo{year}{2009}).

\bibitem[{\citenamefont{Xu et~al.}(2010)\citenamefont{Xu, Engl, Jerison,
  Wallenstein, Hyland, Wilen, and Dufresne}}]{xu.2010}
\bibinfo{author}{\bibfnamefont{Y.}~\bibnamefont{Xu}},
  \bibinfo{author}{\bibfnamefont{W.~C.} \bibnamefont{Engl}},
  \bibinfo{author}{\bibfnamefont{E.~R.} \bibnamefont{Jerison}},
  \bibinfo{author}{\bibfnamefont{K.~J.} \bibnamefont{Wallenstein}},
  \bibinfo{author}{\bibfnamefont{C.}~\bibnamefont{Hyland}},
  \bibinfo{author}{\bibfnamefont{L.~A.} \bibnamefont{Wilen}}, \bibnamefont{and}
  \bibinfo{author}{\bibfnamefont{E.~R.} \bibnamefont{Dufresne}},
  \bibinfo{journal}{Proc. Nat. Acad. Sci.} \textbf{\bibinfo{volume}{107}},
  \bibinfo{pages}{14964} (\bibinfo{year}{2010}).

\bibitem[{\citenamefont{Crocker and Grier}(1996)}]{crocker.1996}
\bibinfo{author}{\bibfnamefont{J.~C.} \bibnamefont{Crocker}} \bibnamefont{and}
  \bibinfo{author}{\bibfnamefont{D.~G.} \bibnamefont{Grier}},
  \bibinfo{journal}{Journal of Colloid and Interface Science}
  \textbf{\bibinfo{volume}{179}}, \bibinfo{pages}{298} (\bibinfo{year}{1996}).

\bibitem[{\citenamefont{Lu et~al.}(2007)\citenamefont{Lu, Sims, Oki, Macarthur,
  and Weitz}}]{lu.2007}
\bibinfo{author}{\bibfnamefont{P.~J.} \bibnamefont{Lu}},
  \bibinfo{author}{\bibfnamefont{P.~A.} \bibnamefont{Sims}},
  \bibinfo{author}{\bibfnamefont{H.}~\bibnamefont{Oki}},
  \bibinfo{author}{\bibfnamefont{J.~B.} \bibnamefont{Macarthur}},
  \bibnamefont{and} \bibinfo{author}{\bibfnamefont{D.~A.} \bibnamefont{Weitz}},
  \bibinfo{journal}{Optics Express} \textbf{\bibinfo{volume}{15}},
  \bibinfo{pages}{8702} (\bibinfo{year}{2007}).

\bibitem[{\citenamefont{Landau and Lifshitz}(1986)}]{landau}
\bibinfo{author}{\bibfnamefont{L.~D.} \bibnamefont{Landau}} \bibnamefont{and}
  \bibinfo{author}{\bibfnamefont{E.~M.} \bibnamefont{Lifshitz}},
  \emph{\bibinfo{title}{Theory of Elasticity}}, vol.~\bibinfo{volume}{7} of
  \emph{\bibinfo{series}{Course of Theoretical Physics}}
  (\bibinfo{publisher}{Butterworth Heinemann}, \bibinfo{address}{Oxford},
  \bibinfo{year}{1986}), \bibinfo{edition}{3rd} ed.

\end{thebibliography}

\end{document}